\documentclass[aps,prb,preprint,showpacs,groupedaddress]{revtex4}

\usepackage{amsmath}
\usepackage{amssymb}
\usepackage{graphicx}
\usepackage{dcolumn}
\usepackage{bm}
\usepackage{subfigure}

\begin{document}

\title{Zero-temperature equation of state of solid $^{4}$He at low and high pressures}

\author{C. Cazorla$^{1,2}$ and J. Boronat$^{3}$}
\affiliation{$^{1}$London Centre for Nanotechnology, UCL, London WC1H OAH, UK \\ 
$^{2}$Department of Physics and Astronomy, UCL, London WC1E 6BT, UK \\
$^{3}$Departament de F\'isica i Enginyeria Nuclear, UPC, Campus Nord B4-B5, Barcelona E-08034, Spain}
\

\begin{abstract}
We study the zero-temperature equation of state (EOS) of solid $^{4}$He in the hexagonal closed packet (hcp) phase
over the $0 - 57$ GPa pressure range by means of the Diffusion Monte Carlo (DMC) method and the semi-empirical Aziz pair potential HFD-B(HE).
In the low pressure regime (P $\sim 0 - 1$ GPa) we assess excellent agreement with experiments and we give an 
accurate description of the atomic kinetic energy, Lindemann ratio and Debye temperature over a wide range of molar volumes
($22 - 6$ cm$^{3}$/mol). However, on moving to higher pressures our calculated $P - V$ curve presents an increasingly steeper
slope which ultimately provides differences within $\sim$ 40$\%$ with respect to measurements. 
In order to account for many-body interactions arising in the crystal with compression which are not reproduced by our model,
we perform additional electronic density-functional theory (DFT) calculations for correcting the computed DMC energies in a perturbative way.
We explore both generalized gradient and local density approximations (GGA and LDA, respectively) for the electronic exchange-correlation potential.
By proceeding in this manner, we show that discrepancies with respect to high pressure data are reduced to $5 - 10 \%$ with few 
computational extra cost. Further comparison between our calculated EOSs and \emph{ab initio} curves deduced for the perfect crystal 
and corrected for the zero-point motion of the atoms enforces the reliability of our approach.
\end{abstract}

\pacs{67.80.-s,61.50.Ah,61.50.Ks}

\maketitle

\section{Introduction}
\label{sec:introduction}

The physics of helium at low temperatures is among the most intriguing and intensively studied topics in condensed matter science.
Despite of being a rare gas element with one of the simplest possible
electronic structures, helium constitutes a fundamental system which is challenging for the test and development 
of methods based on quantum theory. Because of its light atomic mass and weak interatomic interaction,
helium is the only system that remains liquid under its own vapor pressure ($P = 0$) at zero temperature.
Below 2.17 K, liquid $^{4}$He features superfluidity and Bose-Einstein condensation, two
striking and inherent quantum effects.    
With an external pressure of $\sim$ 25 bar, the fluid at $T = 0$ crystallizes into the hexagonal closed packet
structure (hcp), which remains the stable phase of solid helium at $T \neq 0$ and high pressures, made the 
exception of an fcc loop along melting in between $15 - 285$~K.~\cite{driessen86,loubeyre93,zha04} 

Solid helium is manifestly a quantum crystal. In the regime of ultralow temperatures (few mK) this system 
possesses extraordinarily large atomic kinetic energy ($E_{k} \sim$ 24 K) and Lindemann ratio ($\gamma \sim$ 0.26),
 and likewise anharmonic effects on it are of relevance for predicting and understanding its thermodynamic properties.~\cite{glyde94}
Further testimony about the uniqueness of this solid is posed by the long-standing controversy sparked by recent experimental
findings about whether perfect crystalline $^{4}$He may exhibit superfluid-like behavior and Bose-Einstein
condensation (supersolid).~\cite{kim04,rittner06,profkofev05,ceperley04,galli05} 
From a technological side, solid helium also has some relevance since it is considered as the best 
quasi-hydrostatic medium hence modern technologies based on it have emerged and induced considerable progress
in the field of high-pressure experiments.~\cite{takemura01,bell81,downs96,takemura04}

In this paper, we study the zero-temperature equation of state of bulk solid $^{4}$He in the hcp phase over a wide 
pressure range ($0 - 57$~GPa) with the Diffusion Monte Carlo (DMC) method and the HFD-B(HE) Aziz pair potential
(hereafter referred to AzizII),~\cite{aziz87} 
and additionally with electronic density functional theory (DFT) to account for many-body interactions arising in the system with increasing pressure.  
In all the work we differentiate between two pressure regimes, namely low pressure ($0 \leq P \leq 1$ GPa)
and high pressure ($1 < P \leq 57$~GPa).
Quantum Monte Carlo (QMC) methods have proved among the most accurate and reliable tools for solving
quantum many-body problems associated to condensed matter systems.~\cite{hammond94,anderson99,guardiola98,ceperley95}
In particular, the DMC method is a zero-temperature
approach which yields exact estimation (only subject to statistical uncertainty) of the ground-state
energy and related properties of many-boson interacting systems.~\cite{kalos66,kalos70,ceperley79}
During the last few decades this and other Monte Carlo techniques (mainly, the variational Monte Carlo -VMC- and Path Integral Monte Carlo -PIMC-
methods) have been fruitfully applied to the study of noble gases and light elements and compounds like He, Ne, H, D, LiH and LiD in
homogeneous and inhomogeneous phases and both in bulk and in reduced
 dimensionalities.~\cite{sola06,boronat04,gordillo03,grau02,drummond06,cazorla05,cazorla04}
The great capability of the DMC method is related to the existence of accurate interatomic potentials, which are expressed in the form 
of many-body expansions, and are tuned to reproduce empirical and/or theoretical data. Interatomic 
potentials are of precious value because provide computational affordability by allowing one to model atoms as
interacting points (thus avoiding direct treatment of the electronic degrees of freedom of the system), 
 and also simplified understanding of the system under study. 
In the case of helium, the semi empirical pair-potential HFD-HE2 proposed by Aziz \emph{et al.}~\cite{aziz79} more than twenty years ago
 has allowed for quite precise reproduction of the energetic and structural properties of the liquid and solid phases
near equilibrium.~\cite{kalos81,pandharipande83} In this work, we use a newer version of this potential, namely the  HFD-B(HE) one,~\cite{aziz87}
which has demonstrated excellent performance in the description of the liquid~\cite{boronat94} but heretofore has
not been tested in the crystal upon high pressure.

\begin{figure}[t]
\centerline{
        \includegraphics[width=0.8\linewidth]{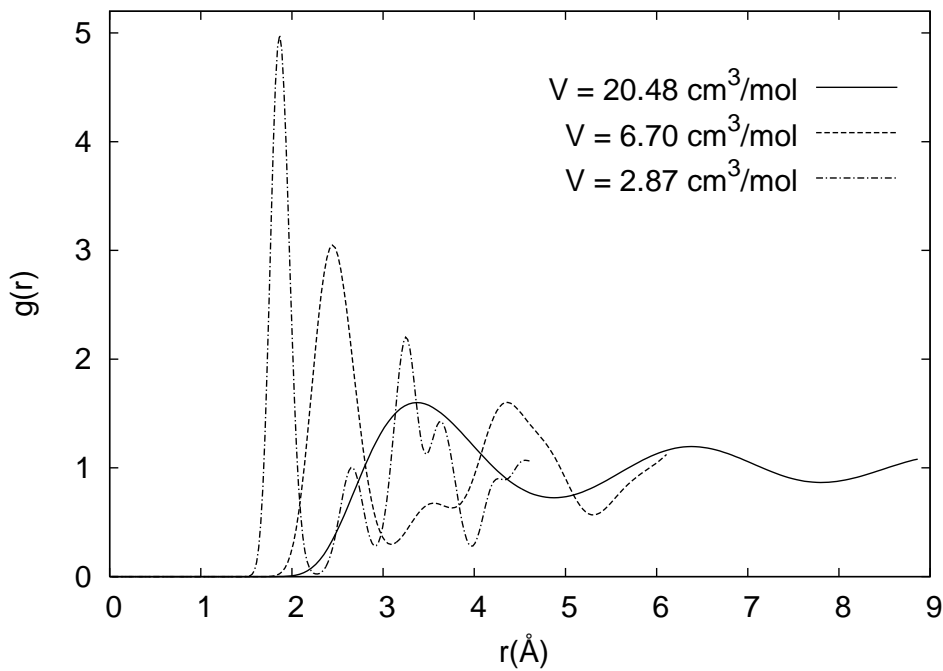}}%
        \caption{ Radial pair distribution function of solid $^4$He at different molar volumes as computed with
		  DMC. Curves are terminated at half the length of the simulation box (containing 180 particles in each case).} 
\label{fig:gr}
\end{figure}

Anticipating some of the outcomes we are to present, excellent agreement between our results for EOS
and experiments is assessed in the low pressure regime for volumes ranging from $V = 21.30$ cm$^{3}$/mol to $V = 8.50$ cm$^{3}$/mol;
however, discrepancies start to develop at smaller volumes ($P > 0.65$ GPa). Within the low pressure regime, we provide exact estimation 
within some statistical error of the kinetic energy per atom, Lindemann ratio and Debye temperature of the system by means of the pure
estimator technique.~\cite{liu74,reynolds86,casulleras95}  
In the high pressure regime, however, our equation of state systematic and increasingly overestimates the pressure.
Discrepancies with respect to measurements amount to $\sim 10\%$ at $P = 1$ GPa and to $\sim 40\%$ at $P = 57$ GPa.
Previous PIMC work on the EOS of solid $^{4}$He at ambient temperature ($T = 300$ K) and performed with akin model pair potentials
arrived to similar disagreements.~\cite{boninsegni01,herrero06} 
With the aim of analyzing the possible causes of this large disagreement we first examine the influence of finite size effects  
in our results. Indeed, finite size effects become larger by increasing pressure because cut-off distances involved in the calculation
of the atomic interactions within the system are continously reduced (generally these are chosen as half the length of the simulation box).
Accordingly, the radial pair distribution function for crystals, $g(r)$, emerges progressively less smooth with compression. Therefore, 
customary corrections devised for dealing with finite size effects which are based on simple approximations for $g(r)$, might introduce
appreciable deviations in the results (see Fig.~\ref{fig:gr}).
Because of these effects, we have regarded as essential to quote the energy tails accounting for the finite size corrections 
by means of two different approaches:
(i) considering $g(r)\simeq 1$ beyond a certain cut-off distance and then integrating the simplified analytic expressions for the tails, and (ii),
computing the variational Monte Carlo energy of progressively enlarged systems and then estimating the energy of the corresponding 
infinite system by means of extrapolation to $N \to \infty$.
Certainly, approach (ii) results computationally more demanding than (i) but also more accurate, and 
we find a pressure difference of $\sim 5$ GPa between both resulting EOSs at the smallest studied volume ($V = 2.50$ cm$^{3}$/mol).
Nevertheless, this discrepancy by itself does not explain the large disagreement between our results and high-pressure data.
In consequence, we turn our main concern to the characterization of the interatomic interactions.

It is well-known that the structural and electronic properties of the atomic and molecular systems may experience important
arrangements by effect of pressure.~\cite{cohen74,cohen84,christensen06,dewaele03} 
Upon compression, overlappings between the electronic clouds within the system are promoted hence further correlations   
among the atoms (angular forces) emerge so as to lower their energies. 
In the case of solid $^{4}$He, it has been suggested and tested within the Self Consistent Phonon formalism that three-body exchange
interactions become significant with increasing density.~\cite{loubeyre87} 
In Ref.~\onlinecite{boninsegni01}, Chang and Boninsegni included three-body effects into their high-pressure PIMC calculations
performed with pair potentials, by computing the energy of several three-body interaction models over sets of configurations 
generated in their simulations (that is, perturbatively).
In doing this, their agreement with experiments did not improve quantitatively, thus they suggested 
that higher order many-body contributions to the energy had to be considered.
More recently, Herrero~\cite{herrero06} has adopted a similar but computationally more demanding approach to that of Chang and Boninsegni in
which three-body interactions are explicitly included into the model Hamiltonian. For the case of a rescaled Bruch-McGee
three-body interaction, Herrero's work provides very notable agreement with experiments up to pressures of $\sim 52$ GPa and at room temperature.  

In the present work, we propose a perturbative approach for quoting the many-body interactions happening within highly compressed
solid $^{4}$He which are not accounted for by any atomic pair potential, and without increasing the computational
cost significantly.
Essentially, this consists in performing \emph{ab initio} density functional calculations over sets of atomic configurations
independently drawn from DMC simulations; subsequently, the energies previously computed with DMC are corrected according to the
average difference between the \emph{ab initio} interaction energy of the all-electron-ion system and the pair-potential energy. 
In this way, many-body interactions of order two and higher are included perturbatively into the EOS without
requiring from the knowledge of any additional two-, three-, four-, and so on, body interatomic potential. 
We show that proceeding in this manner the agreement with high pressure experimental data is at the  $\sim 5-10 \%$ level
with relatively few computational extra cost.
Truly, the approach that here we present for helium can be extended to the study of other quantum crystals
upon high pressure for which accurate pair potentials are available.

The remainder of this paper is as follows. Section \ref{sec:methods} describes the methods that we have
employed, the treatment of finite size effects and gives the technical details.
In Section \ref{sec:results}, we present our results for the ground state of solid $^4$He at low and high pressure and yield
further comparison with first-principles based calculations.
Finally, in Section \ref{sec:discussion} we analyze the pros and cons of the proposed perturbative scheme and conclude with
the final remarks.

\section{Approach and Methods}
\label{sec:methods}

\subsection{Diffusion Monte Carlo}
\label{subsec:DMC}

 DMC is a ground-state method which provides the exact energy within statistical errors  
of many-boson interacting systems of interest.~\cite{hammond94,guardiola98,ceperley79} 
This technique is based on a short-time approximation for the Green's function corresponding 
to the imaginary time-dependent Schr${\rm \ddot{o}}$dinger equation, which is solved 
up to a certain order of accuracy within an infinitesimal interval $\Delta \tau$. 
Despite this method is algorithmically simpler than domain Green's function Monte Carlo,~\cite{ceperley79,kalos74} it presents some
 $\left(\Delta \tau\right)^{n}$
bias coming from the factorization of the imaginary time propagator $e^{-\frac{\Delta\tau}{\hbar}{\rm H}}$.
Our implementation of DMC is quadratic,~\cite{chin90} hence the control of the time-step bias is efficiently controlled since 
the required $\Delta\tau \to 0$ extrapolation is nearly eliminated by choosing a sufficiently small time step.
The Hamiltonian, ${\rm H}$, describing our system is 
\begin{equation}
\label{eq:hamiltonian}
{\rm H} = - \frac{\hbar^{2}}{2m_{\rm He}} \sum_{i=1}^{N} \nabla^{2}_{i} +
       \sum_{i<j}^{N} V_{2}^{AzizII}(r_{ij})~,
\end{equation}  
where $m_{\rm He}$ is the mass of a $^4$He atom, $r_{ij}$ the distance between atoms composing an $i$,$j$ pair
and $V_{2}^{AzizII}(r_{ij})$ the HFD-B(HE) Aziz interaction.~\cite{aziz87}
The corresponding Schr${\rm \ddot{o}}$dinger equation in imaginary time ($it \equiv \tau$), 
\begin{equation}
\label{eq:schrodinger}
-\hbar\frac{\partial \Psi({\bf R},\tau)}{\partial \tau}= \left({\rm H}-E\right)\Psi({\bf R},\tau) 
\end{equation}
with $E$ an arbitrary constant, can be formally solved by expanding the
solution $\Psi({\bf R}, \tau)$ in the basis set of the energy eigenfunctions $\{\phi_{n}\}$. It turns out that
$\Psi({\bf R}, \tau)$ tends to the ground state wave function $\phi_{0}$ of the system for an infinite imaginary time as well as the
expected value of the Hamiltonian tends to the ground-state value $E_{0}$. 
The hermiticity of the Hamiltonian guarantees the equality   
\begin{equation}
\label{eq:groundstate1}
E_{0} = \frac{\left<\phi_{0}|{\rm H}|\phi_{0}\right>}{\left<\phi_{0}|\phi_{0}\right>}=
\frac{\left<\phi_{0}|{\rm H}|\psi_{T}\right>}{\left<\phi_{0}|\psi_{T}\right>} \quad , 
\end{equation}
where $\psi_{T}$ is a convenient trial wave function which depends on the atomic coordinates of the system
${\bf R}\equiv \{ {\bf r_{1}}, {\bf r_{2}},...,{\bf r_{N}} \}$.
Consequently, the ground-state energy of the system can be computed by calculating the integral
\begin{equation}
\label{eq:integral}
E_{DMC}= \lim_{\tau \to\infty} \int_{V} E_{L}\left({\bf R}\right) f\left({\bf R},\tau\right) d{\bf R} = E_{0} \quad ,
\end{equation} 
where $f\left({\bf R},\tau\right)=\Psi\left({\bf R},\tau\right)\psi_{T}\left({\bf R}\right)$ (assuming it is normalized),
and $E_{L}\left({\bf R}\right)$ is the local energy defined as
$E_{L}({\bf R})= {\rm H}\psi_{T}\left({\bf R}\right)/\psi_{T}\left({\bf R}\right)$. 
The introduction of $\psi_{T}\left({\bf R}\right)$ in $f\left({\bf R},\tau\right)$ is known as importance sampling and   
it certainly improves the way in which integral (\ref{eq:integral}) is computed 
(for instance, by imposing $\psi_{T}\left({\bf R}\right)=0$ when $r_{ij}$ is smaller than the core distance
of the interatomic interaction).

In this work, the trial wave function adopted for importance sampling corresponds to the extensively tested Nosanow-Jastrow
model~\cite{nosanow64,hansen68,hansen69}
\begin{equation}
\psi_{\rm NJ}\left({\bf r_{1}},{\bf r_{2}},...,{\bf r_{N}}\right) =
\prod_{i\neq j}^{N} {\rm f_{2}}(r_{ij}) \prod_{i=1}^{N}{\rm g_{1}}(|{\bf r_{i}}-
{\bf R_{i}}|) \quad ,
\label{eq:nosanojashe}
\end{equation}
with ${\rm f_{2}}(r) = e^{-\frac{1}{2}\left(\frac{b}{r}\right)^{5}}$ and ${\rm g_{1}}(r) = e^{-\frac{1}{2}ar^{2}}$ where $a$ and $b$ are variational 
parameters.
This model is composed of two-body correlation functions ${\rm f_{2}}(r)$ accounting for the two-body correlations induced by $V_{2}(r)$, 
and one-body functions ${\rm g_{1}}(r)$ which localize each particle around a site
of the equilibrium lattice of the crystal as given by the set of vectors $\lbrace {\bf R_{i}}\rbrace$.
Initially, the parameters contained in $\psi_{\rm NJ}$ are optimized by means of variational Monte Carlo at some
molar volume near equilibrium; however, as we have explored the system over a wide range of volumes we have repeated this
procedure at some other selected points along the EOS. For instance, the optimized value of the parameters $a$ and $b$ at the molar volume
$20.48$~cm$^{3}$ is $1.12$ and $0.87$~\AA$^{-2}$, respectively, while at $4.02$~cm$^{3}$ results $1.15$ and $3.06$~\AA$^{-2}$~.   
The parameters of the simulations, namely, the number of particles, $N$, critical population of walkers, 
$n_{w}$, and time step, $\Delta\tau$, have been adjusted to eliminate any residual bias coming from them; their
respective values are 180, 400 and 2.7$\cdot$10$^{-4}$ K$^{-1}$. The parameter $\Delta\tau$ has been
reduced progressively with increasing density in order to provide numerical stability.

\subsection{Finite size corrections} 
\label{subsec:corr}

\begin{figure}[t]
\centerline{
        \includegraphics[width=0.8\linewidth]{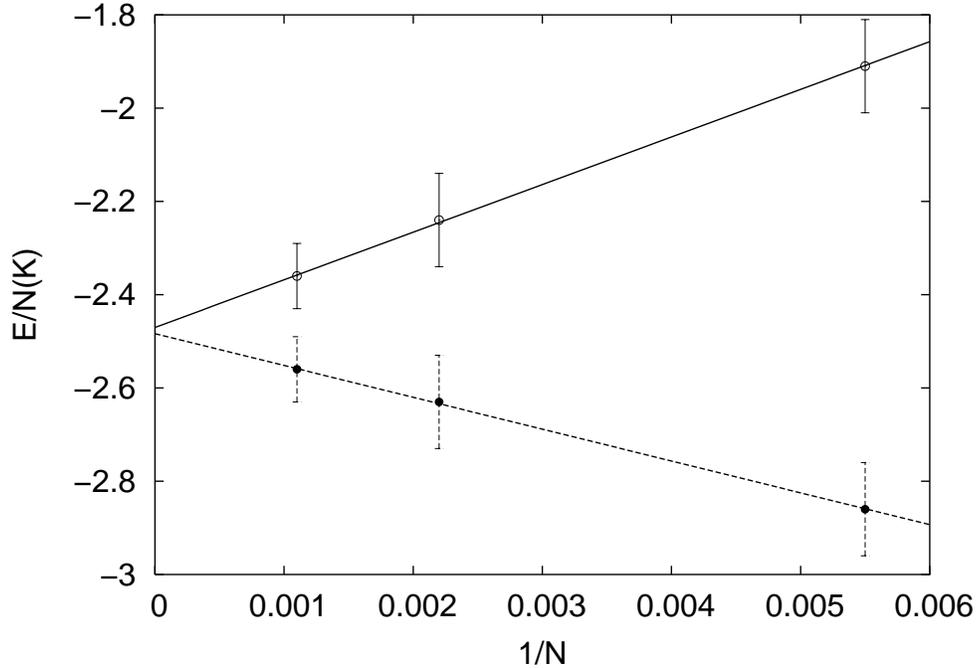}}%
        \caption{ Variational energy per particle in solid $^4$He at $V = 21.35$ cm$^3$/mol as a function of $1/N$.
		  Filled circles correspond to total energy assuming STC energy tail corrections while the empty
		  ones correspond to the total energy deduced directly from the simulation; both respective linear fits
                  are coincident in the limit $N \to \infty$  
	          }
\label{fig:fscorr}
\end{figure}

The description of an infinite system of interacting particles is obtained in practice 
through the simulation of a finite number of particles enclosed within a box.
The difference between the scale of the real and simulated systems can be overcome by
enlarging the size of the simulated system so much as possible and applying periodic boundary
conditions to it.~\cite{allen89} Even so, several corrections to the energies quoted directly from the
simulation must be done if correlations of longer range are present.
Certainly, these corrections arise from the fact that the maximum distance involving  
correlations in the simulation coincides with the length-scale of the particle container.   
The expressions for the potential and kinetic energy corrections $\Delta V ^{tail}$ and
$\Delta T ^{tail}$, assuming a certain cut-off length R$_{max}$ for the computation of the correlations (generally chosen
as half the length of the simulation box), are~: 
\begin{eqnarray} 
\label{eq:epotcorrhe}
\Delta V ^{tail} = 2\pi N \rho \int_{R_{max}}^{\infty} g(r) V_{2}(r) r^{2} dr 
\end{eqnarray}    
\begin{eqnarray} 
\label{eq:ekincorrhe}
\Delta T ^{tail} = -4\pi N D \rho \int_{R_{max}}^{\infty} g(r) \nabla^{2}\ln{{\rm f_{2}}(r)} r^{2} dr ~,
\end{eqnarray}    
where $N$, $D=\hbar^{2}/2m$ and $\rho$ are the number, diffusion constant and density of particles, and 
 $g(r)$, $V_{2}(r)$ and ${\rm f_{2}(r)}$ the radial pair distribution function, pair potential and 
two-body correlation function entering the trial wave function, respectively.
In the case of liquids, $g(r)$ can be well-approximated to unity  
in equations (\ref{eq:epotcorrhe}) and (\ref{eq:ekincorrhe}), and consequently, $\Delta V ^{tail}$ and 
$\Delta T ^{tail}$ turn out to be analytically accessible (standard tail correction -STC-).
 Nevertheless, in the case of solids such approximation could result rather inaccurate owing that the
pattern of the radial distribution function is still oscillating beyond the cut-off distance (see Fig.~\ref{fig:gr}).
In view of these facts and in order for the attained description of solid $^{4}$He to be as precise as possible,  
we have estimated $\Delta V ^{tail}$ and $\Delta T ^{tail}$ also by means of VMC (variational tail correction -VTC-) through the relation
\begin{equation}
\label{eq:vtc}
\Delta E ^{tail} = \Delta T ^{tail} +  \Delta V ^{tail} =
 E_{VMC}^{\infty} - E_{VMC}^{N} ~,  
\end{equation}
where the superscripts in the energies refer to number of particles, $N$ is the number of particles used in the DMC simulations and 
$E_{VMC} \equiv \left< \psi_{T}|{\rm H}|\psi_{T} \right> / \left< \psi_{T}|\psi_{T}\right>$.
The limit $N \to \infty$, equivalent to $R_{max} \to \infty$ in equations (\ref{eq:epotcorrhe}) and (\ref{eq:ekincorrhe}), is reached
through successive enlargements of the simulation box at fixed density (up to 900 particles) and further linear extrapolation to
infinite volume. Indeed, this procedure results computationally affordable within VMC but not within 
 DMC. In Fig.~\ref{fig:fscorr}, we shown the asymptotic agreement between standard and variational energy tail
corrections for infinite solid $^{4}$He ($1/N \to 0$) within VMC.

\subsection{\emph{Ab initio} calculations and perturbative approach} 
\label{subsec:DFT}

Density Functional Theory (DFT) is a first-principles quantum approach which has allowed for accurate and reliable
knowledge of a great deal of materials and systems with exceptional computational affordability. 
A comprehensive description of DFT methods as applied to the modeling of condensed matter is given in recent
books and reviews.~\cite{martin04,kohanoff06}
In DFT, the \emph{ab initio} free energy of an atomic system, given the positions and charges of its nuclei, is expressed
as a functional of the electronic density, $n({\bf r})$, as follows:  
\begin{eqnarray}
E[n({\bf r})] = T[n({\bf r})] + \frac{1}{2}\int\int\frac{n\left({\bf r}\right)n\left(\bf {r'}\right)}{\mid {\bf r}-{\bf r'}\mid}d{\bf r}d{\bf r'} + \sum_{I}^{N} Z_{I} \int\frac{n\left({\bf r} \right)}{\mid {\bf R_{I}}-{\bf r} \mid} d{\bf r} +  \nonumber \\
E_{xc}[n({\bf r})] + \sum_{I<J}^{N} \frac{Z_{I}Z_{J}}{\mid {\bf R_{I}}-{\bf R_{J}} \mid}  , 
\label{eq:dftfunct}
\end{eqnarray}
where $T[n({\bf r})]$ is the electronic kinetic energy, $Z_{I}$ and ${\bf R_{I}}$ the atomic number and position of atom $I$, respectively,
and $E_{xc}[n({\bf r})]$ the electronic exchange-correlation energy (we have imposed $1/4\pi\epsilon_{0}$ and $e$ $\equiv 1$). 
The other terms in Eq.(\ref{eq:dftfunct}) account for the Coulomb interactions between electrons, electrons and nuclei and nuclei. 
The Hohenberg-Kohn theorem states that the density $n_{0}(\bf r)$ which minimizes the  functional $E[n({\bf r})]$
corresponds to the true ground-state density of the system (thus  $E_{0}(\lbrace{\bf R}\rbrace)=E[n_{0}(\bf r)]$)
and that this optimal solution is unique.
It is demonstrated that DFT is an exact electronic ground-state method, whereas the electronic exchange-correlation functional
is not known for most of the systems. Consequently, some approximations for it must be introduced in the calculations.
The most widely used models for $E_{xc}$ are the local density approximation (LDA) and the generalized
gradient approximation (GGA), which have been parameterized by different groups. In this work, we use both Ceperley-Alder (CA) version
of LDA~\cite{ceperley80} and Perdew-Wang (PW91) of GGA,~\cite{perdew92} since \emph{a priori} one can not discern confidently which 
is going to result more reliable for the study. 
A completely independent issue from the choice of  $E_{xc}$ is the implementation of DFT that is used. This mainly concerns 
the way in which the electron orbitals are represented. Here, we use the projector augmented wave (PAW) framework 
developed by Bl${\rm \ddot{o}}$ch~\cite{blochl94} and as implemented in the VASP program.~\cite{kresse93,kresse94,kresse96}

The perturbative approach that we propose for correcting the DMC energies obtained with the 
pair potential $V_{2}^{AzizII}$, consists in averaging the quantity $\Delta E = E_{0}(\lbrace{\bf R}\rbrace) - \sum_{i<j} V_{2}^{AzizII}(R_{ij})$
over sets of configurations drawn independently from the DMC simulations. According to this, the corrected energies result
\begin{equation}
E'_{DMC} \equiv E_{DMC} +  \langle \Delta E\rangle_{DMC} ~.
\label{eq:aicorr}
\end{equation}
The many-body correction $\langle \Delta E\rangle_{DMC}$ includes two-, three- and so on many-body 
contributions to the total energy as can be seen by invoking a many-body expansion of the \emph{ab initio}
ground state energy  
\begin{equation}
\Delta E \equiv E_{0}(\lbrace{\bf R}\rbrace) - \sum_{i<j} V_{2}^{AzizII}(R_{ij}) = \sum_{i<j}^{N} V_{2}(R_{ij}) - V_{2}^{AzizII}(R_{ij}) +  \sum_{i<j<k}^{N}V_{3}(R_{ij},R_{ik},R_{jk}) + \cdots ~.
\label{eq:manybodyexp}
\end{equation}
We note that the family of vectors $\lbrace{\bf R}\rbrace$ here refer to the positions of the atoms (nuclei) and not to the sites
of the perfect crystalline lattice.
It turns out that all the many-body terms composing $\Delta E$ are evaluated for any arrangament of the 
atoms as generated according to the Hamiltonian in Eq.(\ref{eq:hamiltonian}), and included into the total energy
in a perturbative manner.
Certainly, our many-body approach is not exact; 
firstly, it is noted that the full quantum Hamiltonian of the system expressed within the Born-Oppenheimer approximation (FQ-BO) might
be written down as ${\rm H}_{FQ-BO} = \hat{T}_{ion} + E_{0}$, where $\hat{T}_{ion}$ corresponds to the kinetic energy of the nuclei, and that
Eq.(\ref{eq:hamiltonian}) results a simplification of it, needless to be said, extremely accurate at low and moderate pressures.
Nevertheless, using the DMC method for solving the ground-state of 
${\rm H}_{FQ-BO}$ remains a future goal because of the numerous intricacies
encountered in the treatment of the electronic degrees of freedom (i.e. choice of the trial wavefunction and sign
problem) and large computational cost involved.
Therefore, instead of abording the full quantum problem straightaway, we have opted for a simplified but affordable strategy:
add and substract $V_{2}^{AzizII}$ into ${\rm H}_{FQ-BO}$, solve exactly the part of the Hamiltonian embodying most of 
the two-body interactions and account for the rest by means of first-order perturbation theory. 
 
The \emph{ab initio} calculations required for the computation of $\langle \Delta E\rangle_{DMC}$ have been performed on supercells containing
180 particles and with $2\times 2\times 2$ Monkhorst-Pack ${\bf k}$-sampling of the Brillouin zone~\cite{monkhorst76} and cut-off energy 478.0 eV;
these settings ensure energy convergence to better than 0.5 K per atom.
On the other side, our criterion for the convergence of correction $\langle \Delta E\rangle_{DMC}$ relies 
on the measure of its fluctuation, $\delta \Delta E ^{2} \equiv (\Delta E - \langle \Delta E\rangle)^{2}$,
over the same collection of DMC configurations than used for the average $\langle \cdot \rangle_{DMC}$.
Given a molar volume, we have requested $\sqrt{\langle\delta \Delta E ^{2}\rangle}_{DMC}$ to
be less than 1 K per atom, that is approximately $0.1\%$ the total DMC energy obtained at large densities.
In the process of drawing atomic configurations from the DMC runs,
we have imposed the only constraint $|E_{L}\left(\lbrace{\bf R}\rbrace\right)-\langle E_{L}\rangle|< \frac{1}{3}|\langle E_{L}\rangle|$,
 where $E_{L}\left(\lbrace{\bf R}\rbrace\right)$ is the local energy of the considered configuration and $\langle E_{L}\rangle$ the mean
energy calculated over the population of walkers to which it corresponds.
We have proceeded so for avoiding spurious configurations on the averages which otherwise are   
rejected within few steps in the DMC sampling.  
The number of atomic configurations required for the convergence of $\langle \Delta E\rangle_{DMC}$ has proved 
smaller than initially expected in all the studied cases: about 15-25 were enough. This rapid convergence
of the fluctuations $\sqrt{\langle \delta \Delta E ^{2}\rangle}_{DMC}$ reveals that despite two-body interactions by themselves are not
sufficient to attain reliable description of very dense solid $^{4}$He they are still of leading relevance on it.     

\section{Results}
\label{sec:results}

\subsection{Low pressure regime}
\label{subsec:lowpress}

\begin{figure}[t]
\centerline{
        \includegraphics[width=0.8\linewidth]{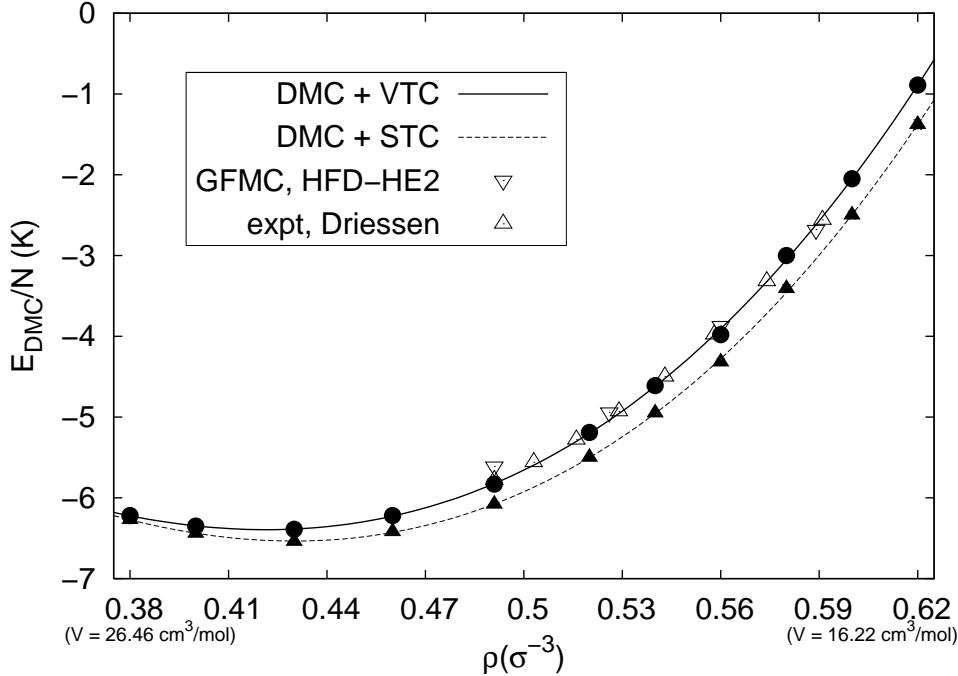}}%
        \caption{Total energy per particle of solid $^{4}$He at low pressures as function of density (expressed in units of $\sigma = 2.556$~\AA)
                 computed with DMC and the HFD-B(HE) Aziz interaction. Results are obtained with VTC ($\bullet$) and STC ($\blacktriangle$)
		 and compared to experimental data of Ref.~\onlinecite{driessen86} ($\vartriangle$) and previous GFMC calculations ($\triangledown$)
		 performed with the HFD-HE2 Aziz potential found in Ref.~\onlinecite{kalos81}. Error bars are smaller than the depicted symbols.}
\label{fig:energyeq}
\end{figure}

The EOSs of solid $^4$He has been obtained by fitting a fourth-order polynomial to the DMC energies and 
subsequently performing the derivation respect to volume,
\begin{equation}
P(V) = -\frac{\partial E}{\partial V} = 2\left(\frac{V_{0}}{V^{2}}\right)\left(\frac{V_{0}}{V} - 1\right)
\left( a + 2b\left( \frac{V_{0}}{V} - 1\right)^{2} \right) ~,
\label{eq:eos}
\end{equation}
where $V_{0}$ is directly the equilibrium volume of the system and $a$, $b$ constants.
In Fig.~\ref{fig:energyeq},  we compare  the DMC energies at volumes close to equilibrium ($P(V) \sim 0$),  obtained
with both STC and VTC, with  experimental data.~\cite{driessen86} As one observes there, excellent
agreement between measurements and VTC results is provided, however, differences with respect to STC results are quoted
in an almost constant upwards shift of $\sim 0.30$ K at positive pressure. 
Despite these discrepancies will practically vanish when expliciting both VTC and STC EOSs because of the energy derivative involved
(as it will be shown shortly), it is noted that for other magnitudes which explicitly depend on the internal energy, as for example
the enthalpy or freezing and melting densities, VTC and STC lead to different results.
In the same figure, we also display previous theoretical calculations performed with Green's function Monte Carlo (GFMC) and the HFD-HE2 Aziz pair
potential.~\cite{kalos81} The GFMC points perfectly coincide with our results obtained with VTC, however, they disagree with the STC ones
in the same manner experimental points do. Since GFMC and DMC are exact ground-state methods, energy differences between both approaches
should be due only to the model interaction. Assuming that the treatment of finite size effects adopted in Ref.~\onlinecite{kalos81} 
corresponds to the commonly 
used STC one, we may just conclude that both HFD-HE2 and HFD-B(HE) interatomic potentials are likely to produce equivalent $P-V$ curves
at low pressures (likewise VTC and STC lead to practically identical EOSs) but not so total energies and other 
directly related properties. 
 
\begin{figure}[t]
\centerline{
        \includegraphics[width=0.8\linewidth]{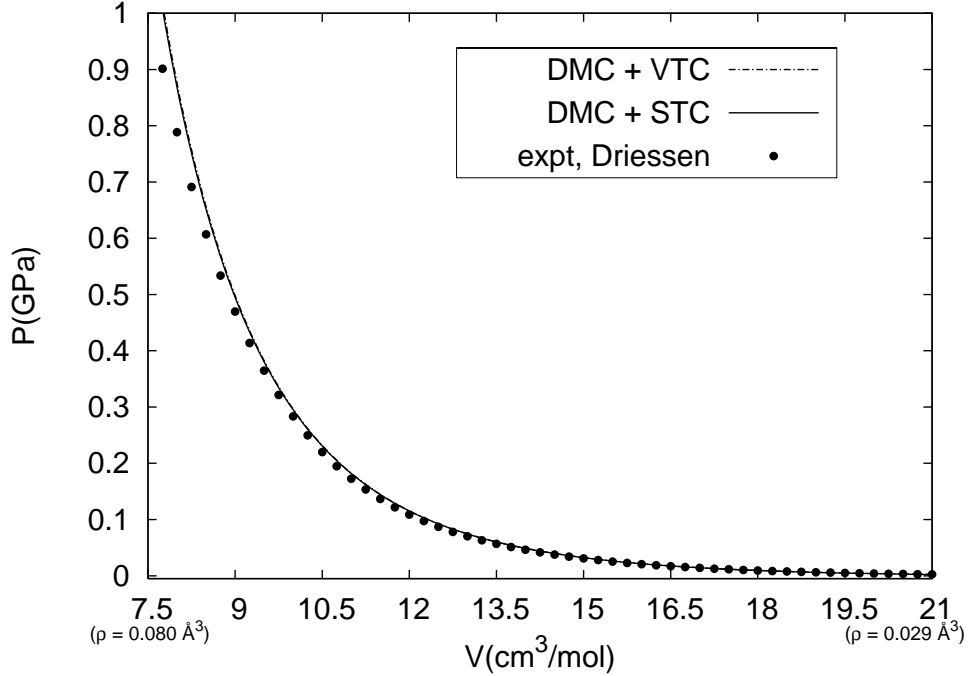}}%
        \caption{EOS of solid $^{4}$He as computed with DMC and HFD-HE2 interatomic potential in the pressure range
		 $0\leq P \leq 1$~GPa. VTC and STC lead to identical curves within the statistical uncertainty and
		 experimental data of Ref.~\onlinecite{driessen86} (points) is provided for comparison.   
	           }
\label{fig:varpresslow}
\end{figure}

Fig.~\ref{fig:varpresslow} reports our results of the EOS of solid $^{4}$He at $T = 0$ in the volume range $21.0 \leq V \leq 7.5$ cm$^{3}$/mol 
($0 \le P \le 1.0$ GPa). Curves obtained with VTC and STC are coincident as we anticipated in the previous paragraph.
The parameters of the fit (\ref{eq:eos}) also displayed in Fig.~\ref{fig:varpresslow} (we provide the 
one obtained with VTC) are $a = 9.11(6)$~K, $b = 16.93(15)$~K and $V_{0} = 25.04(4)$~cm$^{3}$/mol (uncertainties are shown within parentheses).
A glance at the plot reveals an excellent agreement between our results and experiments at low pressures, however, discrepancies become
progressively larger as we move towards volumes smaller than $8.5$ cm$^{3}$/mol ($P \sim 0.65$ GPa). 
(For instance, at $V = 7.76$ cm$^{3}$/mol our prediction of pressure overestimates the experimental 
value within $\sim 10\%$.)
It is worth noticing that the worsening of our curve roughly coincides with the interval in which the potential energy of the system becomes positive
(see Table I). This fact indicates that the repulsive part of the HFD-B(HE) potential is probably too stiff. 
In the next subsection we will extensively deal with the shortcomings derived from the adopted model interaction,
however now we continue with the description of other atomic magnitudes of interest that we have obtained at low pressures.

\begin{figure}[t]
\centerline{
        \includegraphics[width=0.8\linewidth]{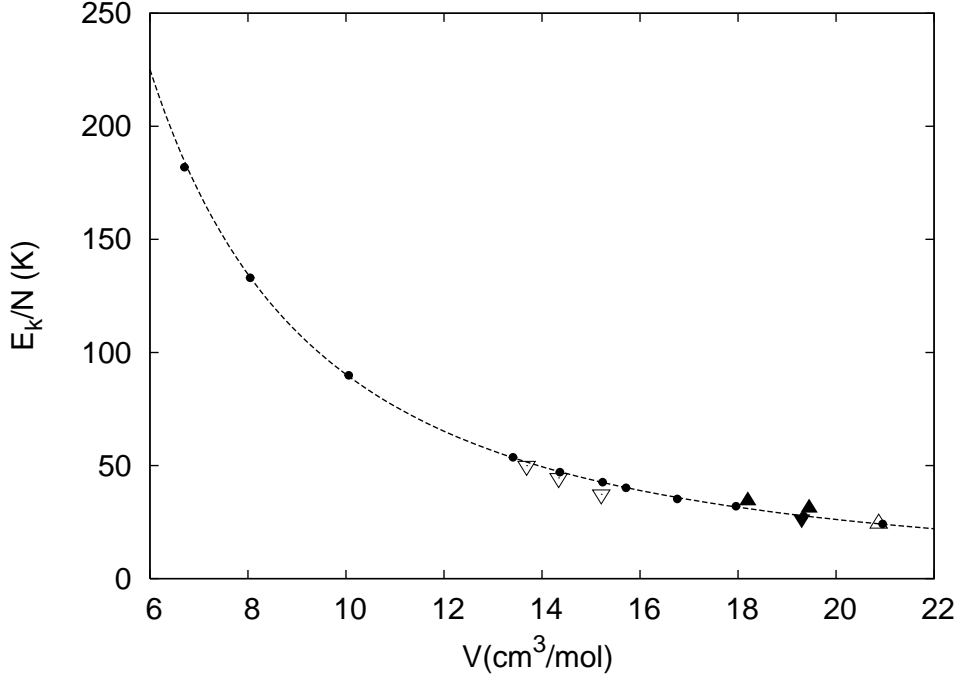}}%
        \caption{ Kinetic energy per particle of solid $^{4}$He, $E_{k}/N$, as computed with DMC and HFD-HE2 potential ($\bullet$
                  and guide-to-eye $---$). Experimental data found in Refs.~\onlinecite{hilleke84} ($\blacktriangle$), ~\onlinecite{diallo04}
		  ($\vartriangle $), ~\onlinecite{adams07} ($\blacktriangledown$) and ~\onlinecite{celli98} ($\triangledown$) are provided
		  for comparison. Error bars are smaller than the depicted symbols. }
\label{fig:kineticlowp}
\end{figure}

The zero-temperature atomic kinetic energy of solid $^{4}$He, $E_{k}$, is an important (and challenging) quantity to measure and  
compute since it evidences the singular quantum nature of this crystal.
It is well-known that the zero-point energy of solid helium is comparable in magnitude to its potential energy (cohesive energy),
$E_{p}$, and that the ratio between these two energies gives a qualitative idea about the relevance of
anharmonic effects in the system (the larger $E_{k}/E_{p}$ is, the larger anharmonic contributions would result).~\cite{glyde94,moleko85}   
From a computational side, \emph{exact} estimation of the expected ground state values of operators which do not commute
with the Hamiltonian, as for instance the potential and kinetic energy operators, may be provided within the DMC scenario by
means of the pure estimator technique.~\cite{liu74,reynolds86,casulleras95} In practice, this technique involves the introduction of additional weight
factors into the customary DMC sampling which retain memory of the configurational replication processes occurring along the simulation.
In order for our evaluation of the zero-temperature kinetic energy of solid $^{4}$He to be as reliable as possible, we first determine the exact
potential energy of the system by means of the pure estimator method and then we subtract it to the total energy
(we note that within DMC the estimator of the kinetic energy is slightly biased by the choice of the trial wavefunction).
In Fig.~\ref{fig:kineticlowp} we display our results for $E_{k}$ and compare them to low temperature data provided by several
authors:~\cite{hilleke84,diallo04,adams07,celli98} the overall agreement between them is excellent.
In particular, we note the perfect agreement of our calculated value $E_{k} = 24.24~(5)$~K at $V = 20.87$~cm$^{3}$/mol with
the very recent neutron scattering measurement of Diallo {\it et al.},~\cite{diallo04} $E_{k}^{expt} = 24.25~(30)$~K, performed at the same volume.
In Table I, we enclose DMC results for the total, kinetic and potential energies of solid $^{4}$He including STC at some
selected volumes within the interval $22.60$-$8.0$ cm$^{3}$/mol. In Fig.~\ref{fig:kineticlowp}, however, we have not refrained from including
a further point at volume $6.70$ cm$^{3}$/mol for which we have shown that the description of the system attained with the model interaction seems to be
not fully reliable. It should be mentioned that the treatment of finite size effects adopted in the calculations has little effect on the final 
values of $E_{k}$ since the largest contribution to the total energy tail correction stems from the interatomic interactions.  

\begin{figure}[t]
\centering
       { \includegraphics[width=0.45\linewidth]{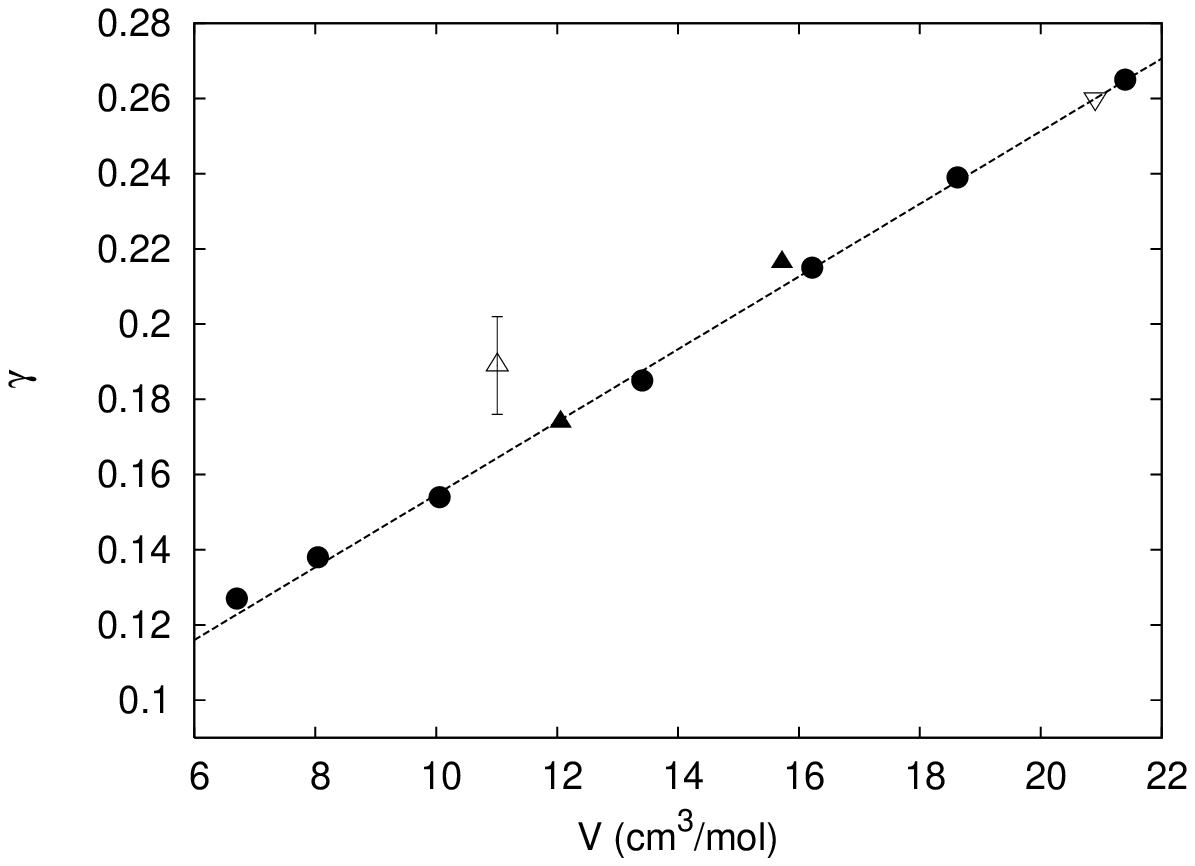} }%
       { \includegraphics[width=0.45\linewidth]{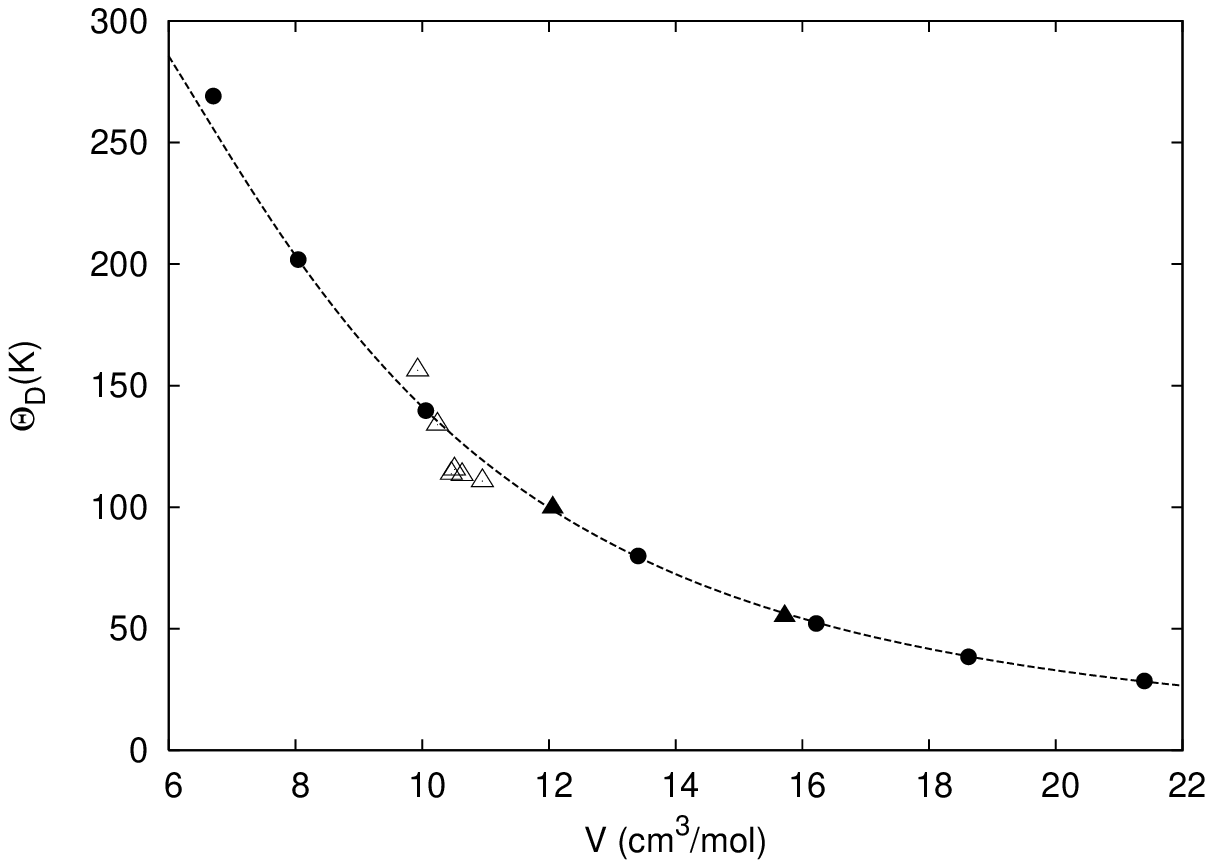} }%
 \vspace{0.0cm}
 \caption{ \emph{Left}: Lindemann ratio of solid $^{4}$He as function of volume. Our results are $\bullet$ and line $0.058 + 
			 0.0097~ V$ as guide-to-eye, and experimental data of Ref.~\onlinecite{stassis78} ($\blacktriangle$), ~\onlinecite{venkataraman03}
			 ($\vartriangle$) and ~\onlinecite{burns97} ($\triangledown$) are shown for comparison.
	   \emph{Right}: Debye Temperature of solid $^{4}$He at $T = 0$ as function of volume. Our results are $\bullet$ and line $---$,
			 and experimental data of Ref.~\onlinecite{venkataraman03} ($\vartriangle$) and ~\onlinecite{stassis78} ($\blacktriangle$)
			 are shown for comparison.}
\label{fig:rlindebye}
\end{figure}

Another quantity of interest in the study of quantum solids is the atomic mean squared displacement, $\langle {\rm {\bf u}}^{2}\rangle$~, which
is directly measured in x-ray diffraction experiments. 
In connection to this magnitude, the Lindemann ratio is defined as $\gamma = \sqrt{\langle {\rm {\bf u}}^{2}\rangle}/d$, 
where $d$ is the distance between nearest neighbors in the perfect crystalline lattice. As we pointed out in the Introduction, 
the zero-temperature Lindemann ratio of solid $^{4}$He is uncommonly large (even if compared to other distinguished 
quantum solids like for example H$_{2}$ which possesses $\gamma \sim 0.18$) as consequence of its light atomic mass and weak interatomic interaction. 
Using the pure estimator technique, we have studied the dependence of $\gamma$ with volume over the range $22.6$-$8.0$~cm$^{3}$/mol.
We have depicted our results for $\gamma$ in Fig.~\ref{fig:rlindebye} and compared them to experimental data of different authors, and again
the overall agreement between them is remarkable. Once $\langle {\rm {\bf u}}^{2}\rangle$ is known, the Debye temperature of the system at $T = 0$, 
$\Theta _{D}$, is deduced straightforwardly through the relation $\Theta_{D} = 9\hbar^{2}/ 4m_{\rm He} \langle {\rm {\bf u}}^{2}\rangle$~.
We have fitted our results for $\Theta_{D}$ with the relation 
\begin{equation}
\Theta_{D} = \exp{\left( \sum_{i=0}^{3} c_{i}x^{i} \right)}~,
\label{eq:fitdebye}
\end{equation}
where $x \equiv \ln{\left(V/V_{D}\right)}$ and which has been used previously to reproduce the density dependence of
the phonon frequencies in solid H$_{2}$ and $^{4}$He as well.~\cite{driessen84,driessen86} Our optimal coefficients for expression (\ref{eq:fitdebye}) 
plotted in Fig.~\ref{fig:rlindebye} are: $V_{D} = 22.6166$ cm$^{3}$/mol, $c_{0} = 3.21655$~, $c_{1} = -2.23859$~, 
$c_{2} = 0.122057$ and $c_{3} = 0.319911$~. 
(Additional point at $V = 6.70$ cm$^{3}$/mol in Fig.~\ref{fig:rlindebye}, has not been used in the fit.)

\begin{table}
\begin{center}
\begin{tabular}{c c c c}
\hline
\hline
$\qquad  V ({\rm cm}^{3}/{\rm mol}) \qquad $   &   $\qquad  E_{DMC}/N ({\rm K}) \qquad $  &   $\qquad  E_{k}/N ({\rm K}) \qquad $  &   $\qquad  E_{p}/N ({\rm K}) \qquad $  \\
\hline
$ 22.60  $  &   $ -6.51(2)  $  &   $ 21.36(6) $  &   $ -27.87(6) $   \\ 
$ 20.95  $  &   $ -6.22(2)  $  &   $ 24.20(6) $  &   $ -30.42(6) $   \\ 
$ 19.34  $  &   $ -5.50(2)  $  &   $ 27.63(6) $  &   $ -33.13(6) $   \\ 
$ 17.96  $  &   $ -4.32(2)  $  &   $ 32.01(6) $  &   $ -36.33(6) $   \\ 
$ 16.76  $  &   $ -2.50(2)  $  &   $ 35.24(6) $  &   $ -37.74(6) $   \\ 
$ 15.24  $  &   $  1.63(2)  $  &   $ 42.63(6) $  &   $ -41.00(6) $   \\ 
$ 14.37  $  &   $  5.25(3)  $  &   $ 47.09(8) $  &   $ -41.84(7) $   \\ 
$ 13.41  $  &   $  11.11(5) $  &   $ 53.66(9) $  &   $ -42.55(8) $   \\ 
$ 10.06  $  &   $  68.80(5) $  &   $ 89.90(9) $  &   $ -21.10(8) $   \\ 
$ 8.04   $  &   $ 192.45(5) $  &   $133.00(9) $  &   $  59.45(8) $   \\ 
\hline
\hline
\end{tabular}
\end{center}
\caption{ Total, kinetic and potential energies per particle of solid $^{4}$He including STC ($E_{DMC}$, $E_{k}$ and $E_p$, respectively)
 as computed with DMC and the HFD-B(HE) Aziz potential. Figures within parentheses account for the statistical errors.}
\label{tab:energies}
\end{table}

\begin{figure}
\centering
       { \includegraphics[width=0.8\linewidth]{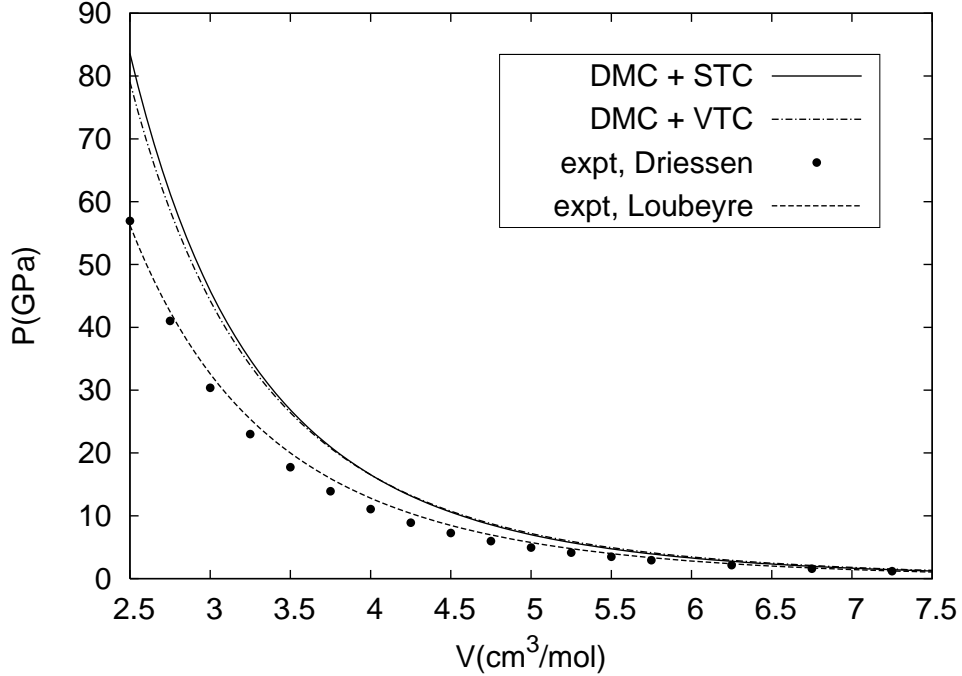} }%
 \vspace{0.0cm}
 \caption{ Equations of state of solid $^{4}$He over the $0-57$ GPa pressure range as computed with DMC and HFD-B(HE) interaction.
	   Experimental data of Ref.~\onlinecite{driessen86} ($\bullet$) and ~\onlinecite{loubeyre93} (dashed line) are included for comparison.}
\label{fig:dmchighpress}
\end{figure}

\begin{figure}
\centering
       { \includegraphics[width=0.8\linewidth]{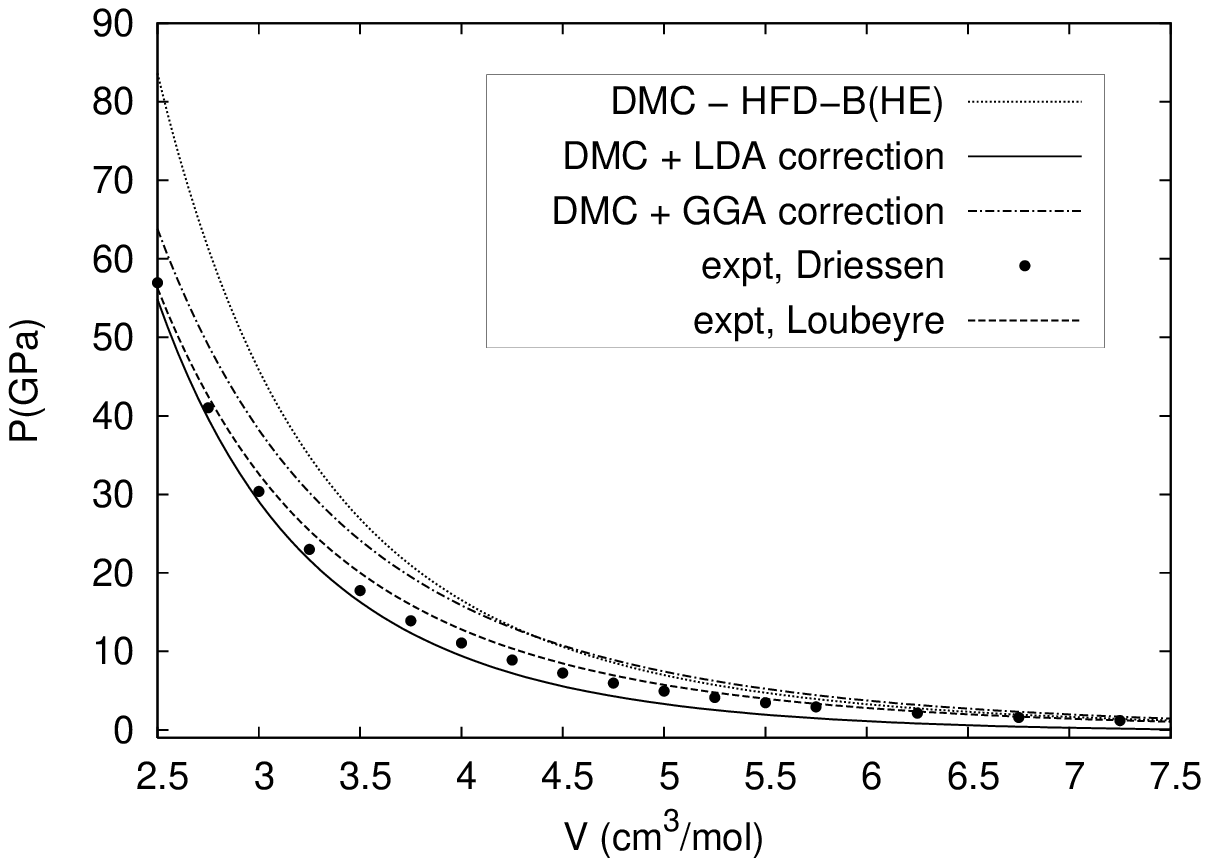} }%
 \vspace{0.0cm}
 \caption{Zero-temperature EOS of solid $^{4}$He as computed with DMC and HFD-B(HE) pair potential and considering 
	  perturbative many-body corrections to the energy (solid and dashed-dotted lines mean LDA and GGA corrections, respectively).
	  Experimental data of Ref.~\onlinecite{driessen86} ($\bullet$) and ~\onlinecite{loubeyre93} (dashed line) are enclosed for comparison.  
  }
\label{fig:perturbation}
\end{figure}

\subsection{High pressure regime}
\label{subsec:highpress}

As we have already illustrated in Sec.\ref{subsec:lowpress}, the pair potential HFD-B(HE) performs excellently in the description
of solid $^{4}$He up to volumes of $8.5$ cm$^{3}$/mol, however, it monotonically fails to reproduce its EOS as the density is
increased beyond this point.     
In Fig.~\ref{fig:dmchighpress}, we explicit the differences between measurements of Ref.~\onlinecite{driessen86} and ~\onlinecite{loubeyre93}
and our calculations performed with STC and VTC for finite size effects and over the pressure range $0-57$ GPa.
The pressure difference between the EOSs obtained with VTC and STC at the highest studied density amounts to $\sim 5$ GPa,
however this quantity results very small when compared to the discrepancy of both with experiments which is 
$\sim 40\%$ of the experimental value. This discrepancy is in overall agreement with the microscopic calculations of 
Refs.~\onlinecite{herrero06} and ~\onlinecite{boninsegni01}.

Our results and other found in the literature,~\cite{boninsegni01,herrero06} pose the need for considering higher order many-body 
effects present on dense $^4$He instead of going in the search of improved pair potential models. 
As we pointed out in the Introduction, many authors have made efforts for elucidating the
relevance of three-body and higher order (up to six-body) effects on the EOS of solid $^{4}$He with assorted degree of accuracy
and success.~\cite{boninsegni01,herrero06,tian06}
In this section, we present the $P-V$ curves calculated within our proposed scheme for correcting the DMC energies obtained with
pair potentials as described in Sec.~\ref{subsec:DFT}.
Just as we have explained there, all the many-body interactions not accounted for by $V_{2}^{AzizII}$ are 
computed with \emph{ab initio} DFT and included into the total energy in a perturbative way, without requiring
from the knowledge of additional two-, three- and/or higher order many-body interaction models. 

The fits to our results displayed in the $P-V$ figures of this and next subsections have been performed
with the Vinet relation~\cite{vinet87} 
\begin{equation}
P(V) = 3 B_{0} \left(1 - \left(\frac{V}{V_{0}}\right)^{\frac{1}{3}}\right)\left(\frac{V_{0}}{V}\right)^{\frac{2}{3}}\exp{\left[ \frac{3}{2}
\left(B_{0}^{'} - 1\right) \left(1 - \left(\frac{V}{V_{0}}\right)^{\frac{1}{3}}\right)  \right]}~,
\label{eq:vinet}
\end{equation}
where $V_{0}$, $B_{0}$ and $B_{0}^{'}$ are the equilibrium volume, equilibrium isothermal bulk modulus ($B\equiv -V \partial P/\partial V$)
and equilibrium $\partial B/\partial P$, respectively. The experimental values of these parameters as provided by Ref.~\onlinecite{loubeyre93}  
are $V_{0}^{expt} = 13.72$ cm$^{3}$/mol, $B_{0}^{expt} = 0.225$ GPa and $B_{0}^{'expt} = 7.35$~. We have also enclosed 
data points of Ref.~\onlinecite{driessen86} in the plots for additional comparison with our estimations.
The improvement of our EOS when considering many-body effects computed with the proposed perturbative approach is substantial
(see Fig.~\ref{fig:perturbation}). For example, within LDA we obtain $P = 54.83$ GPa at volume $2.5$ cm$^{3}$/mol
which is quite close to the experimental value $56.94$ GPa and far below the non-corrected DMC result of $83.60$ GPa.
The parameters of the fit corresponding to this case are $V_{0}^{LDA} = 7.77$ cm$^{3}$/mol, $B_{0}^{LDA} = 1.884$ GPa 
and $B_{0}^{'LDA} = 6.66$~, which as it is observed in Fig.~\ref{fig:perturbation} leads to constant underestimation of pressure
within few Gpa respect to experimental data along the whole depicted range.
On the contrary, the EOS obtained with GGA provides a notable description of the system near the equilibrium and
few GPa above the experimental values at high pressures.
Putting this into figures, $V_{0}^{GGA} = 12.93$ cm$^{3}$/mol, $B_{0}^{GGA} = 0.510$ GPa
and $B_{0}^{'GGA} = 6.53$~, which in fact result closer to the experimental values of Ref.~\onlinecite{loubeyre93} than the LDA ones.
It is worth mentioning that the observed tendency of GGA (LDA) for overestimating (underestimating) pressure in our results is a well-known outcome in
the field of \emph{ab initio} simulations.

 Table II yields the values of the DMC energy and perturbative corrections for solid $^{4}$He
at some selected volumes. Separately, we have shifted all the LDA and GGA corrections 
a same amount for providing null contributions at the largest enclosed volume so as to facilitate the comparison between them.
Certainly, this can be done without any loss of generality since the zero of the LDA and GGA \emph{ab initio} energies and that 
of the HFD-H(BE) interaction do not coincide and we are essentially interested in the pressure. Two main conclusions can be extracted from  
the values $\langle \Delta E\rangle_{DMC}$ in Table II: (i) corrections performed with LDA decrease
monotonically with compression, not so with GGA, and (ii) GGA corrections are smaller in absolute value 
than the LDA ones. Since the proposed approach for correcting the DMC energies obtained with pair potential
is perturbative, the conclusion (ii) concedes more reliability to the results obtained with GGA than with LDA.
Indeed, a conclusive answer about whether LDA figures are or not too large would be best provided by second order perturbative theory,
 however this is out from our scope. In the next subsection, we will comment again on this issue by supplying further comparison 
between results presented here and others obtained by means of \emph{ab initio} procedures.     

\begin{table}
\begin{center}
\begin{tabular}{c c c c}
\hline
\hline
$\quad  V ({\rm cm}^{3}/{\rm mol}) \quad $   &   $\quad  E_{DMC}/N ({\rm K}) \quad $  &   $\quad \langle\Delta E\rangle_{DMC}^{LDA}/N ({\rm K}) \quad $  &   $\quad \langle \Delta E \rangle_{DMC}^{GGA}/N ({\rm K}) \quad $  \\
\hline
$ 10.06  $  &   $ 68.80(5)     $  &   $  0.00(75)    $  &  $ 0.00(14)     $   \\ 
$  6.70  $  &   $ 404.55(5)    $  &   $ -352.55(88)  $  &  $ 72.06(33)    $   \\ 
$  5.03  $  &   $ 1163.54(8)   $  &   $ -813.08(55)  $  &  $ 200.43(39)   $   \\ 
$  4.02  $  &   $ 2444.11(12)  $  &   $ -1407.99(50) $  &  $ 232.38(35)   $   \\ 
$  3.35  $  &   $ 4294.67(15)  $  &   $ -2165.61(61) $  &  $ 43.53(50)    $   \\ 
$  2.87  $  &   $ 6728.33(38)  $  &   $ -3113.77(67) $  &  $ -389.66(51)  $   \\ 
$  2.51  $  &   $ 9742.06(49)  $  &   $ -4263.34(98) $  &  $ -1055.16(98) $   \\ 
\hline
\hline
\end{tabular}
\end{center}
\caption{Calculated DMC energies and corrections $\langle\Delta E\rangle_{DMC}$ per particle for solid $^{4}$He at some selected volumes.
         Within the parentheses are the statistical uncertainties, which in the case of the corrections correspond to 
	 $\sqrt{\langle \delta \Delta E^{2} \rangle}_{DMC}/N$ (we note that $(98)\equiv \pm 0.98$ and $(5)\equiv \pm 0.05$).}
\label{tab:perturcorr}
\end{table}

\begin{table}
\begin{center}
\begin{tabular}{c c c c c c c}
\hline
\hline
$  $   &   $\quad Expt. \quad$  &  $\quad DMC\quad$  &   $\quad LDA \quad$  &  $\quad GGA \quad$ & $\quad LDA(vdW)\quad$ & $\quad GGA(vdW)\quad $ \\
\hline
$V_{0} ({\rm cm}^{3}/{\rm mol})$  &   $ 13.72  $  & $ 20.16 $ &  $ 7.77   $  &   $ 12.93  $  &  $ 7.06  $ & $ 11.35  $  \\ 
$B_{0} ({\rm GPa})             $  &   $ 0.225  $  & $ 0.018 $ &  $ 1.884  $  &   $ 0.510  $  &  $ 3.072 $ & $ 0.901  $  \\ 
$B_{0}^{'}                     $  &   $ 7.35   $  & $ 9.85  $ &  $ 6.66   $  &   $ 6.53   $  &  $ 6.19  $ & $ 6.13   $   \\ 
$P_{max} ({\rm GPa})           $  &   $ 56.94  $  & $ 83.60 $ &  $ 54.83  $  &   $ 63.79  $  &  $ 52.42 $ & $ 62.00  $   \\ 
\hline
\hline
\end{tabular}
\end{center}
\caption{Parameters of the fits performed with relation (\ref{eq:vinet}) for the resulting EOSs. The headers on this first row correspond
to experimental values of Ref.~\onlinecite{loubeyre93}, DMC calculations with pair potential HFD-B(HE), DMC calculations with many-body corrections
as obtained with LDA, GGA, LDA plus vdW interaction and GGA plus vdW interaction, respectively. $P_{max}$ is the value of the pressure obtained at
the smallest studied volume $2.5$ cm$^{3}$/mol. 
           }
\label{tab:parameterfits}
\end{table}

One known weakness of DFT calculations is that the usual approximations for $E_{xc}$ may fail to capture the essence of the long-range
forces present in the system.~\cite{kohn98,basanta05} In the case of rare gases, 
the van der Waals (vdW) energy, which physically accounts for Coulomb correlations between distant electrons, 
has notorious relevance in the cohesion of the system.
With the aim of estimating the effect of this shortcoming in our corrections, we have added an effective two-body
term accounting for the vdW interactions to the \emph{ab initio} energy $E_{0}$. This term is expressed as 
\begin{equation}
V_{vdW}(R) = f(R)\frac{C_{6}}{R^{6}}~, 
\label{eq:vdW}
\end{equation}
where $C_{6} = -10130.639$ K\AA$^{6}$ and $f(R) = \exp{\left(-\left(D/R - 1 \right)^{2}\right)}$ for $R < D$ but $f(R) = 1$ for $R > D$
with $D = 4.392944$~\AA~ (that is, as given by the HFD-B(HE) interaction), and it has been evaluated over
the same sets of atomic configurations than used for the computation of $\langle \Delta E\rangle_{DMC}$.
Following this receipt, the many-body correction devised for energies $E_{DMC}$ now can be redefined 
as $\Delta E_{vdW} = E_{0}(\lbrace{\bf R}\rbrace) + \sum_{i<j} V_{vdW}(R_{ij}) - \sum_{i<j} V_{2}(R_{ij})$.
In Fig.~\ref{fig:DFTcomp}, we plot the curves obtained with the 
correction $\Delta E_{vdW}$, and Table III summarizes the parameters of all the fits that we have performed (for LDA and GGA corrections
including and not including vdW contributions).
On one hand, a glance at the figure reveals that considering vdW interactions as explained above has in general little effect on the 
results, just a slight and otherwise expected lowering of the $P-V$ curves within few GPa over the whole depicted range.
On the other hand, the equilibrium properties of the system, as given by the parameters of the fits, change appreciably
(see figures enclosed in Table III). This result seems to corroborate the accepted assumption that the effect of long-range interactions
in the EOS of rare-gas solids becomes less important with increasing pressure.~\cite{kim06,dewhurst02,iitaka02,kwon95}

In this subsection, we have not attempted to enclose any result for fcc $^{4}$He in the plots and/or tables since experiments 
indicate that hcp is the only stable phase of solid helium at high pressures ($\sim$~GPa) and low temperatures,
apart from a small fcc loop region around melting between $15$ and $285$~K.~\cite{driessen86,loubeyre93} 
Reassuringly, previous work based on first-principles calculations agrees to regard hcp as the most energetically favourable 
zero-temperature phase of $^{4}$He upon pressures up to $160$~GPa.~\cite{nabi05} 
In spite of this, we have carried out a series of calculations in highly compressed fcc $^{4}$He in order to check 
the predictability of our approach. Essentially, our results show no appreciable energy differences between the two phases
within the numerical uncertainty.
This outcome, however, appears to be not surprising since short-range interactions in helium are of leading importance, and the first and second 
shells of nearest neighbours in the fcc and hcp phases peak at practically indentical distances given a same density.

\subsection{Comparison with \emph{ab initio} based calculations}
\label{subsec:DFTcomp}

\begin{figure}
\centerline{
        \includegraphics[width=0.8\linewidth]{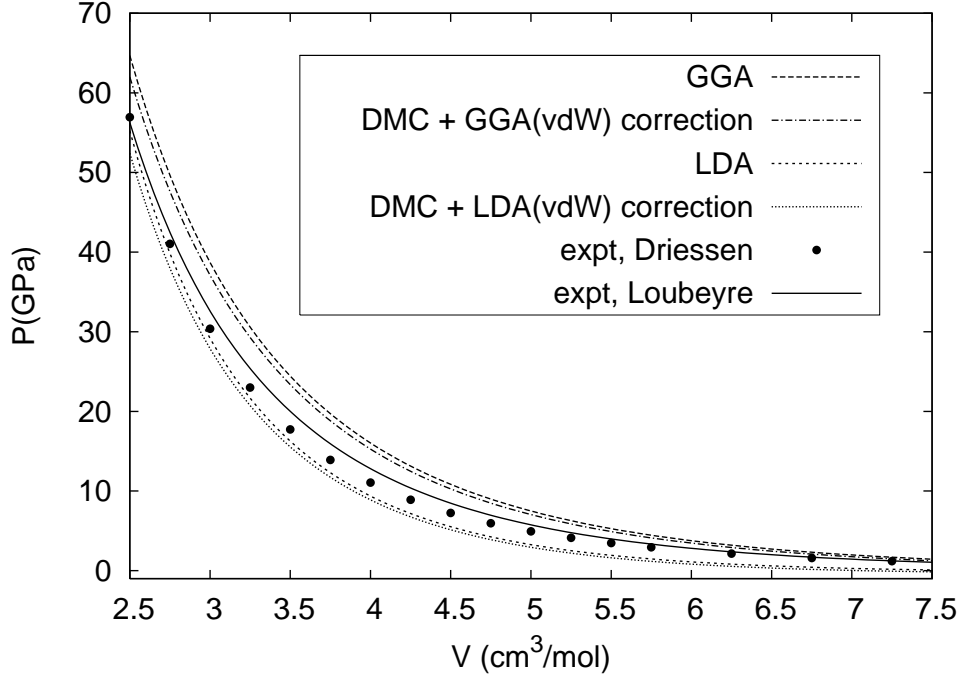}}%
        \caption{EOS of solid $^{4}$He as computed with DFT and corrected for the zero-point motion of the atoms with the
		 Mie-Gruneisen model (GGA and LDA). Curves presented in Sec.(\ref{subsec:highpress}) and experimental data
		 of Ref.~\onlinecite{driessen86} and ~\onlinecite{loubeyre93} are also included for comparison.}
\label{fig:DFTcomp}
\end{figure}

Within the DFT formalism, the zero-temperature energy of a solid is usually written as a sum of two different contributions
\begin{equation}
E_{0}(V) = E_{perf}(V) + E_{vib}(V)~, 
\label{eq:DFTener}
\end{equation}
where $E_{perf}(V)$ is the energy of the perfect crystal (atoms frozen on their sites) and 
$E_{vib}(V) = E_{harm}(V) + E_{anharm}(V)$ accounts for the motion of the atoms and is expressed as a sum of 
harmonic and anharmonic terms. In practice, $E_{perf}$ is obtained with standard DFT calculations and it involves 
affordable computations performed within one unit cell of the perfect crystal (apart from the summations involving periodic boundary
conditions). On the other side, the estimation of $E_{vib}$ requires from some knowledge of the phonon-related properties of the solid of interest.
In the case of heavy-ion crystals, the quasiharmonic approximation in combination with finite displacement methods have allowed
for an accurate description of the phonon frequency spectra.~\cite{alfe01,vocadlo02}
The basic strategy underlying these methods consists in distorting the   
perfect crystal by displacing certain selected atoms slightly from their equilibrium positions and then evaluating the atomic forces arising on the 
system by means of the Hellman-Feynman theorem and DFT. This approach, however, fails to reproduce solid $^{4}$He since it provides negative
(imaginary) phonon frequencies associated to its experimental stable phases at intermediate pressures.~\cite{cazorla07}
Truly, the relevance of anharmonic effects
in solid $^{4}$He makes the computation of its vibrational properties a tedious and complicate task which requires from approaches
going beyond the harmonic and/or quasiharmonic approximations.
It should be noted that within DMC this difficulty is circumvented
since the phononic nature of the studied system is inherently cast into the method,
 hence further partition of the energy into static and vibrational parts is not required.

In order to contrast our results presented in Sec.~\ref{subsec:highpress} with other obtained with \emph{ab initio} based methods,
we have computed the EOS of solid $^{4}$He through the relation (\ref{eq:DFTener}) by evaluating $P_{perf}$ with DFT and 
$P_{vib}$ with the Mie-Gruneisen model~\cite{hemley90}
\begin{equation}
P_{vib}(V) = -\frac{\partial E_{vib}}{\partial V} = \frac{9R\Theta_{D}\gamma_{G}}{8V} ~, 
\end{equation}
where $\Theta_{D}$ is the Debye temperature, $\gamma_{G}$ the Gruneisen parameter which we approximate 
as $\gamma_{G}\equiv -\partial \ln{\omega_{D}}/ \partial \ln{V}$ (with $\hbar \omega_{D} = k_{B}\Theta_{D}$) and $R$
the gas constant. Indeed, we have used for $\Theta_{D}(V)$ the experimental relation provided by Driessen
\emph{et al.}~\cite{driessen86} since, as we have noted previously, the estimation of this or any other vibrational property of 
solid $^{4}$He at $T = 0$ would result not trusty with customary \emph{ab initio} strategies used in the study
of normal (not quantum) solids.  
The resulting $P_{vib}$ increases monotonically with compression and for instance it represents about $15\%$ of the total pressure
of the system at volume $2.5$ cm$^{3}$/mol, thus it must not be neglected.
In Fig.~\ref{fig:DFTcomp}, we show the EOSs obtained with the already explained procedure using both LDA and GGA approximations
for the exchange-correlation energy;
also we include the $P-V$ curves quoted within DMC and corrected for the vdW energy and many-body interactions,
 and experimental data.
As it is observed there, differences between both LDA and GGA perturbationally corrected curves
and their \emph{ab initio} counterparts are quite small. Moreover, these discrepancies
are likely caused by the treatment of the long range interactions (vdW energy) described in Sec.~\ref{subsec:highpress} and
the approximation adopted for $P_{vib}$. This result is stimulating since it 
demonstrates that with the approach presented in Sec.~\ref{subsec:DFT} one can obtain accurate EOSs for dense solid helium, or equivalently for any other light quantum solid,
 in excellent agreement with those which would be obtained by means of first-principles approaches but with the benefit of not requiring
from the computation of the phonon dispersion curves of the crystal or experimental data.

Now we turn our attention to the concern posed over LDA in the previous subsection.
As we noted there, a glance at Table II might lead to the conclusion that the LDA corrections are too large so as to be considered 
perturbative on top of the DMC energies (not so the GGA ones). Indeed, we do not dispose of a fair criterion for accepting or rejecting
corrections in basis to their size, and in our opinion this is the most important shortcoming of our approach.
Nevertheless, appealing to the good accordance between the LDA $P-V$ curve corrected for the atomic zero-point motion and
the DMC one corrected with LDA, we may feel quite confident about the reliability of the latter.

\section{Discussion and Conclusions}
\label{sec:discussion}

In the Introduction we pointed out that Diffusion Monte Carlo is among the best suited methods for studying quantum solids.
In the case of bosonic systems, this method provides the exact ground state energy and related properties
without dependence on the choice of the trial wave function, which otherwise is related to the computational efficiency.
Here, we have proved the excellent performance of DMC with the Aziz pair potential HFD-B(HE) in  
characterizing solid $^{4}$He at low pressures ($P \leq$ 0.65 GPa), by estimating its EOS,  
kinetic energy per particle and Debye temperature at different volumes, and comparing our results with experiments.   
Especial attention has been paid to the extend of finite size effects in our results. To this regard, we conclude that 
customary strategies devised to correct such effects based on the approximation of the pair radial distribution function
to unity beyond certain cut-off distance, may result accurate enough for the derivation of the EOS but not so for the assessment
of other magnitudes like the energy.  

On the other side, solid helium under high compression (as most of the materials) undergoes important arrangements on electronic structure
which lead to the appearance of angular correlations among the atoms.~\cite{dewaele03} This circumstance makes necessary to consider 
not only atomic pair interactions but also higher order many-body ones when investigating this crystal upon high pressure. 
Nevertheless, within DMC the minimal inclusion of three-body interactions on the model Hamiltonian has already the effect of 
drastically slow down the simulations. Furthermore, even in the supposed case computational cost was not a problem, first we should 
know much better than now the analytical form of these many-body interactions or alternatively to be
able to devise them (which actually may result puzzling).
According to this occurrence, \emph{ab initio} methods emerge among the best candidates for quoting such contributions
since they do not rely on potential models and, in general, are computationally affordable. However, fully \emph{ab initio} analysis
of crystals requires from the knowledge of the phonon-related properties, and for the case of solid $^{4}$He and other light quantum solids  
this is by no means a straightforward task.    

In this work, we have presented an approach for the study of dense solid $^{4}$He at zero temperature 
which combines the versatility of the DMC method with the accuracy of \emph{ab initio} calculations.
On one hand, we naturally circumvent the calculation of the vibrational properties of helium thanks to the DMC strategy, and
on the other, we account for the many-body interactions having place on the system by means of DFT. 
 However, the way by which we enclose these many-body contributions to the DMC energy is not exact but perturbative and we do
not dispose of rigorous tests for quoting the errors included on these corrections.
Concerning results, we have yielded the EOSs corrected in this manner within LDA and GGA for the exchange-correlation energy and they have proved
in fairly notable agreement with experiments over the pressure range $0-57$ GPa. Specifically, GGA provides a better description of the 
crystal near equilibrium than LDA.
Further comparison of these curves with EOSs obtained trough DFT and corrected for the atomic zero-point motion by means of an approximate model,
 comes to support the reliability of our approach.   

It should be mentioned that the zero-temperature scheme proposed in this work is also well suited for the study of other light quantum solids
upon high pressure, like $^{3}$He, H$_{2}$, D$_{2}$ and Li, for which accurate pair potentials are devised.
Certainly, a further and promising improvement of the present framework would consist in going beyond the perturbative approach. This could be 
achieved by proper coupling of the DMC and DFT methods, as for instance, by considering the \emph{ab initio} energy of the system within 
the branching weight of the DMC algorithm. Work on this direction is in progress.

\acknowledgments
We acknowledge financial support from DGI (Spain) Grant No. FIS2005-04181 and Generalitat de Catalunya Grant No. 2005GR-00779.


\begin{thebibliography}{99}

\bibitem{driessen86} A. Driessen, E. van der Poll and I. F. Silvera, Phys. Rev. B \textbf{33}, 3269 (1986).
\bibitem{loubeyre93} P. Loubeyre, R. LeToullec, J. P. Pinceaux, H. K. Mao, J. Hu and R. J. Hemley, Phys. Rev. Lett. \textbf{71}, 2272 (1993).
\bibitem{zha04} C-S. Zha, H-K. Mao and R. J. Hemley, Phys. Rev. B \textbf{70}, 174107 (2004).
\bibitem{glyde94} H. R. Glyde in {\it Excitations in Liquid and Solid Helium} (Clarendon Press, Oxford, 1994).
\bibitem{kim04} E. Kim and M. H. W. Chan, Science \textbf{305}, 1941 (2004).
\bibitem{rittner06} A. S. C. Rittner and J. D. Reppy, Phys. Rev. Lett. \textbf{97}, 165301 (2006).
\bibitem{profkofev05} N. Profkof'ev and B. Svistunov, Phys. Rev. Lett. \textbf{94}, 155302 (2005).
\bibitem{ceperley04} D. M. Ceperley and B. Bernu, Phys. Rev. Lett. \textbf{93}, 155303 (2004).
\bibitem{galli05} D. Galli, M. Rossi and L. Reatto, Phys. Rev. B \textbf{71}, 140506 (2005).
\bibitem{takemura01} K. Takemura, J. Appl. Phys. \textbf{89}, 662 (2001).
\bibitem{bell81} P. M. Bell and H. K. Mao, Year Book - Carnegie Inst. Washington \textbf{80}, 404 (1981).
\bibitem{downs96} R. T. Downs, C. S. Zha, T. S. Duffy and L. W. Finger, Am. Mineral. \textbf{81}, 51 (1996).
\bibitem{takemura04} K. Takemura, K. Sato, H. Fujihisa and M. Onoda, Ferroelectrics \textbf{305}, 103 (2004).
\bibitem{aziz87} R. A. Aziz, F. R. W. McCourt and C. C. K. Wong, Mol. Phys. \textbf{61}, 1487 (1987).
\bibitem{hammond94} B. L. Hammond, W. A. Lester Jr. and P. J. Reynolds in {\it Monte Carlo Methods in Ab initio Quantum Chemistry}
                    (World Scientific, 1994).
\bibitem{anderson99} J. B. Anderson, Rev. Comp. Chem. \textbf{13}, 133 (1999).
\bibitem{guardiola98} R. Guardiola in {\it Microscopic Quantum Many-Body Theories and Their Applications} ed. by J. Navarro and
		      A. Polls (Springer, Berlin, 1998).
\bibitem{ceperley95} D. M. Ceperley, Rev. Mod. Phys. \textbf{67}, 279 (1995).  
\bibitem{kalos66} M. H. Kalos, J. Comp. Phys. \textbf{1}, 257 (1966).
\bibitem{kalos70} M. H. Kalos, Phys. Rev. A \textbf{2}, 250 (1970).
\bibitem{ceperley79} D. M. Ceperley and M. H. Kalos in {\it Monte Carlo Methods in Statistical Physics} (Springer, Berlin, 1979).
\bibitem{sola06} E. Sola, J. Casulleras and J. Boronat, Phys. Rev. B \textbf{73}, 092515 (2006).
\bibitem{boronat04} J. Boronat, C. Cazorla, D. Colognesi and M. Zoppi, Phys. Rev. B \textbf{69}, 174302 (2004).
\bibitem{gordillo03} M. C. Gordillo, J. Boronat and J. Casulleras, Phys. Rev. B \textbf{68}, 125421 (2003).
\bibitem{grau02} V. Grau, J. Boronat and J. Casulleras, Phys. Rev. Lett. \textbf{89}, 045301 (2002).
\bibitem{drummond06} N. D. Drummond and R. Needs, Phys. Rev. B \textbf{73}, 024107 (2006).
\bibitem{cazorla05} C. Cazorla and J. Boronat, J. Low Temp. Phys. \textbf{139}, 645 (2005).
\bibitem{cazorla04} C. Cazorla and J. Boronat, J. Low Temp. Phys. \textbf{134}, 43 (2004).
\bibitem{aziz79} R. A. Aziz, V. P. S. Nain, J. S. Carley, W. L. Taylor and G. T. Mc Conville, J. Chem. Phys. \textbf{70}, 4330 (1979).
\bibitem{kalos81} M. H. Kalos, M. A. Lee, P. A. Whitlock and G. V. Chester, Phys. Rev. B \textbf{24}, 115 (1981).
\bibitem{pandharipande83} V. R. Pandharipande, J. G. Zabolitsky, S. C. Pieper, R. B. Wiringa and V. Helmbrehct, Phys. Rev. Lett. \textbf{50},
                          1676 (1983).
\bibitem{boronat94} J. Boronat and J. Casulleras, Phys. Rev. B \textbf{49}, 8920 (1994).
\bibitem{liu74} K. S. Liu, M. H. Kalos and G. V. Chester, Phys. Rev. A \textbf{10}, 303 (1974).
\bibitem{reynolds86} P. J. Reynolds, R. N. Barnett, B. L. Hammond and W. A. Lester Jr., J. Stat. Phys. \textbf{43}, 1017 (1986).
\bibitem{casulleras95} J. Casulleras and J. Boronat, Phys. Rev. B \textbf{52}, 3654 (1995).
\bibitem{boninsegni01} S. Y. Chang and M. Boninsegni, J. Chem. Phys. \textbf{115}, 2629 (2001).
\bibitem{herrero06} C. P Herrero, J. Phys.: Condens. Matter \textbf{18}, 3469 (2006).
\bibitem{cohen74} S. G. Louie and M. L. Cohen, Phys. Rev. B \textbf{10}, 3237 (1974).
\bibitem{cohen84} K. J. Chang and M. L. Cohen, Phys. Rev. B \textbf{30}, 004774 (1984).
\bibitem{christensen06} N. E. Christensen and D. L. Novikov, Phys. Rev. B \textbf{73}, 224508 (2006).
\bibitem{dewaele03} A. Dewaele, J. H. Eggert, P. Loubeyre and R. Le Toullec, Phys. Rev. B \textbf{67}, 094112 (2003).
\bibitem{loubeyre87} P. Loubeyre, Phys. Rev. Lett. \textbf{58}, 1857 (1987).
\bibitem{kalos74} M. H. Kalos, D. Levesque and L. Verlet, Phys. Rev. A \textbf{9}, 2178 (1974).
\bibitem{chin90} S. A. Chin, Phys. Rev. A \textbf{42}, 6991 (1990).
\bibitem{nosanow64} L. H. Nosanow, Phys. Rev. Lett. \textbf{13}, 270 (1964).
\bibitem{hansen68} J. P. Hansen and D. Levesque, Phys. Rev. \textbf{165}, 293 (1968).
\bibitem{hansen69} J. P. Hansen, Phys. Letters \textbf{30A}, 214 (1969).
\bibitem{allen89} M. P. Allen and D. J. Tildesley in {\it Computer Simulation of Liquids} (Oxford University Press, 1989).
\bibitem{martin04} R. M. Martin in {\it Electronic Structure} (Cambridge University Press, 2004).
\bibitem{kohanoff06} J. Kohanoff in {\it Electronic Structure Calculations for Solids and Molecules: Theory and Computational Methods}
                     (Cambridge University Press, 2006).
\bibitem{ceperley80} D. M. Ceperley and B. I. Alder, Phys. Rev. Lett. \textbf{45}, 566 (1980).
\bibitem{perdew92} J. P. Perdew and Y. Wang, Phys. Rev. B \textbf{45}, 13244 (1992).
\bibitem{blochl94} P. E. Bl${\rm \ddot{o}}$chl, Phys. Rev. B \textbf{50}, 17953 (1994).
\bibitem{kresse93} G. Kresse and J. Hafner, Phys. Rev. B \textbf{47}, 558 (1993).
\bibitem{kresse94} G. Kresse and J. Hafner, Phys. Rev. B \textbf{49}, 14251 (1994).
\bibitem{kresse96} G. Kresse and J. Furthm${\rm \ddot{u}}$ller, Comput. Mat. Sci. \textbf{6}, 15 (1996). 
\bibitem{monkhorst76} H. J. Monkhorst and J. D. Pack, Phys. Rev. B \textbf{13}, 5188 (1976).
\bibitem{moleko85} L. K. Moleko and H. R. Glyde, Phys. Rev. Lett. \textbf{54}, 901 (1985).
\bibitem{hilleke84} R. O. Hilleke, P. Chaddah and R. O. Simmons, Phys. Rev. Lett. \textbf{52}, 847 (1984).
\bibitem{diallo04} S. O. Diallo, J. V. Pearce, R. T. Azuah and H. K. Glyde, Phys. Rev. Lett. \textbf{93}, 075301 (2004).
\bibitem{adams07} M. A. Adams, J. Mayers, O. Kirichek and R. B. E. Down, Phys. Rev. Lett. \textbf{98}, 085301 (2007).
\bibitem{celli98} M. Celli, M. Zoppi and J. Mayers, Phys. Rev. B \textbf{58}, 242 (1998).
\bibitem{stassis78} C. Stassis, D. Khatarnian and G. R. Kline, Solid State Comms. \textbf{25}, 531 (1978).
\bibitem{venkataraman03} C. T. Venkataraman and R. O. Simmons, Phys. Rev. B \textbf{68}, 224303 (2003).
\bibitem{burns97} C. A. Burns and E. D. Isaacs, Phys. Rev. B \textbf{55}, 5767 (1997).
\bibitem{driessen84} A. Driessen and I. F. Silvera, J. Low Temp. Phys. \textbf{54}, 361 (1984).
\bibitem{tian06} C. Tian, F. Liu, F. Jing and L. Cai, J. Phys.: Codens. Matt. \textbf{18}, 8103 (2006).
\bibitem{vinet87} P. Vinet, J. R. Smith, J. Ferrante and J. H. Rose, Phys. Rev. B \textbf{35}, 1945 (1987).
\bibitem{kohn98} W. Kohn, Y. Meir and D. E. Makarov, Phys. Rev. Lett. \textbf{80}, 4153 (1998).
\bibitem{basanta05} M. A. Basanta, Y. J. Dappe, J. Ortega and F. Flores, Europhys. Lett. \textbf{70}, 355 (2005).
\bibitem{kim06} E. Kim, M. Nicol, H. Cynn and C. Yoo, Phys. Rev. Lett. \textbf{96}, 035504 (2006).
\bibitem{dewhurst02} J. K. Dewhurst, R. Ahuja, S. Li and B. Johansson, Phys. Rev. Lett. \textbf{88}, 075504 (2002).
\bibitem{iitaka02} T. Iitaka and T. Ebisuzaki, Phys. Rev. B \textbf{65}, 012103 (2002).
\bibitem{kwon95} I. Kwon, L. A. Collins, J. D. Kress and N. Troullier, Phys. Rev. B \textbf{52}, 16165 (1995). 
\bibitem{nabi05} Z. Nabi, L. Vitos, B. Johansson and R. Ahuja, Phys. Rev. B \textbf{72}, 172102 (2005).
\bibitem{alfe01} D. Alf\`{e}, G. D. Price and M. J. Gillan, Phys. Rev. B \textbf{64}, 045123 (2001).
\bibitem{vocadlo02} L. Vo\v{c}adlo and D. Alf\`{e}, Phys. Rev. B \textbf{65}, 214105 (2002).
\bibitem{cazorla07} {\it As it has been checked by the authors}.
\bibitem{hemley90} R. J. Hemley, H. K. Mao, L. W. Finger, A. P. Jephcoat, R. M. Hazen and C. S. Zha, Phys. Rev. B \textbf{42}, 6458 (1990).

\end{thebibliography}
\end{document}